\documentstyle[aps,preprint]{revtex}
\tightenlines
\begin{document}
\title{The charge radius of a D$p$brane}
\author{Michael Gutperle and Vipul Periwal}
\address{Department of Physics,
Princeton University,
Princeton, New Jersey 08544}

\def\dd{\hbox{d}}
\def\tr{\hbox{tr}}\def\Tr{\hbox{Tr}}
\maketitle
\begin{abstract}  Sen has shown that tachyon condensation in
Dbrane--anti-Dbrane configurations can lead to remarkable connections
between string theories.  A consequence of his results
is that there is a minimal value of
the radial co\"ordinate $r_{c} \propto \sqrt{\alpha'}
(gN)^{1/(9-p)}$ such that N units of D$p$brane charge   cannot be
localized to values smaller than $r_{c}.$  At this value of $r_{c}$
the curvature and the gradient of the Ramond-Ramond field strength
are of order $1/\alpha',$ and the vacuum, regarded as a
Dbrane--anti-Dbrane configuration {\it with} a tachyon condensate, is
rendered unstable, leading to a separation of the Dbrane and the
anti-Dbrane.  This value of $r_{c}$ lies in the region intermediate
between the near-horizon regime and the asymptotic regime for Dbrane
classical solutions for small $g.$   This vacuum stability bound on
the curvature can be interpreted as an uncertainty relation for Dbrane
charge.
\end{abstract}
\vskip 1.0cm

\def\al{\alpha}
\def\be{\beta}
\def\la{\lambda}
\def\eps{\epsilon}
\def\sig{\sigma}
\def\la{\lambda}
\def\ga{\gamma}
\def\half{\hbox{$1\over 2$}}
\def\quart{\hbox{$1\over 4$}}
\def\ee#1{\hbox{e}^{{#1}}}
\def\part{\partial}
\def\refe#1{eq.~(\ref{#1})\ }
Sen\cite{sen,sen1} has recently found remarkable exact descriptions of
Dbrane--anti-Dbrane configurations that exhibit crucial connections
between non-BPS states in dual string theories.  Witten\cite{witten}
related
these results to $K$--theory, putting the specific examples found by
Sen\cite{sen,sen1}  into a general framework.  In this paper, we examine
a particular  physical consequence  of these ideas\cite{sen,sen1,witten}.

Dbrane--anti-Dbrane configurations are not supersymmetric by themselves.
Indeed,
they exhibit tachyonic instabilities for short distances\cite{green,banks}.
The particular aspect of Sen's work that will be of interest to us is
that tachyon condensation leads to a stable configuration of
a Dbrane lying on top of an anti-Dbrane\cite{sen,sen1}.  With such a
tachyon condensate, Sen\cite{sen,sen1}\  and Witten\cite{witten}\
have argued that this configuration is equivalent to the vacuum,
hence it is supersymmetric in particular.   This
stable configuration of a Dbrane--anti-Dbrane pair with a tachyon
condensate has the important property that the open string mode
that corresponds to separating the Dbrane from the anti-Dbrane
becomes massive.  Tachyon condensation
leading to stable configurations appeared also in
\cite{others,polchinski,srednicki}.

In \cite{sen1}\ Sen found an exact conformal field theory
description of an anti-Dstring--Dstring pair with a ${\bf Z}_{2}$ Wilson line
on one of the strings.  This turned out, when the tachyon was given an
expectation value, to be a configuration describing a particle, the
strong coupling limit of the
massive Spin(32)/${\bf Z}_{2}$ spinor state of the heterotic string.
In particular, this configuration was shown to possess SO(9)
invariance, which implies that, except at the position of the kink of the
tachyon field produced by the presence of a Wilson line, along the
anti-Dstring--Dstring pair, the configuration is equivalent to
`nothing', as mentioned above.  See also the discussion in Witten\cite{witten}.
Sen\cite{sen1}\ also showed that the mode corresponding to relative
motion of the Dstring--anti-Dstring acquires a mass when the tachyon
condenses.  This mode is a linear combination of the U(1) modes
corresponding to motion of the Dstring and the anti-Dstring, and is not
projected out by the GSO projection since it is a mode that
corresponds to strings stretched from the Dstring to itself, and
strings stretched from the anti-Dstring to itself.

We expect that these features should appear in any such
Dpbrane--anti-Dpbrane
configuration, independent of $p.$  Specifically we assume that
(1) `nothing' is equivalent to a D$p$brane--anti-D$p$brane with
a tachyon condensate, and that
(2) modes corresponding to relative
motion of the D$p$brane--anti-D$p$brane acquire  a finite mass when the
tachyon
condenses.
As indicated in the preceding paragraph, these facts were
explicitly demonstrated by Sen for the Dstring--anti-Dstring
configuration with a Wilson line, but the presence of the Wilson line
does not seem to imply anything other than the formation of a kink in
the tachyon condensate. From a spacetime perspective, it is
difficult to  see how  either
of our assumptions depends on the presence of the Wilson line.
We are extrapolating
from demonstrated facts\cite{sen1}\ about Dstrings to general
D$p$branes.  It could be that new features arise in general that
invalidate our analysis.

In principle, the annulus diagram with the analogue of
Sen's conformal field theory
{\it without} the Wilson line should describe the force between
Dbranes and anti-Dbranes in the presence of a tachyon condensate.
However, we have not succeeded in eliminating the Wilson line from
his construction\cite{sen1}\ so we are forced to rely on a spacetime
perspective for our conclusions.  As we discuss below, we will only
need a `sub--stringy' domain where the forces due to supergraviton
exchange are still small, so this spacetime perspective may be
justified.

An important point:
Our discussion will be phrased in terms of a radial co\"ordinate
$r$ transverse to the branes.  We will be computing gradients in
the radial direction at such values of the radius that covariant
derivatives can be replaced by partial $r$ derivatives to lowest order
in $gN.$  Since the proper distance to the co\"ordinate $r=0$
is infinite for $p<4,$ it is not strictly correct in these cases to speak of
the charge radius, since $r_{c}$ is an infinite proper distance away
from $r=0.$  The actual determination of $r_{c}$ below will rely only
on the magnitude  of the gradient of the force felt by an
anti--Dbrane, and is therefore independent of the proper distance to
$r=0.$

A detailed analysis of the classical size
of Dbranes was given by Barb\'on\cite{jose}\ some time ago.
It is informative
to compare his discussion with that given below.

We consider such a `probe' Dbrane--anti-Dbrane pair {\it with} a tachyon
condensate in the background produced by a source of $N$ Dbranes.
Remark that it
is an abuse of common terminology to talk of this configuration as a
`probe'
because it is equivalent to the vacuum\cite{sen,sen1,witten}---what we
are actually considering is the polarization of the vacuum produced by
the $N$ Dbranes.  Whenever we use the term `probe' in the following, we always
mean the brane--antibrane--tachyon condensate configuration equivalent
to the vacuum.  The
probe Dbrane feels no force due to the source because of its BPS
character, but the probe
anti-Dbrane feels a static attraction towards the source.  Since the
open string mode  corresponding to separating the probe pair is
massive, this attraction merely gives this field in the world-volume
theory of the probe pair an expectation value.  However, because the
source sets up a static force on the
Dbrane--anti-Dbrane probe that
has a non-vanishing gradient, it follows that at a critical value of
the radial co\"ordinate
$r_{c}$ between the probe and the source, the separation mode becomes
massless and the anti-Dbrane can combine with the source.  The
end result is a configuration in which the probe pair has been reduced
to a probe Dbrane, and the source configuration has $N-1$ Dbranes.
Thus, $r_{c}$ measures the critical value of the radial
co\"ordinate below which it is not
possible to localize Dbrane charge.  The  dependence of this
critical value of $r_{c}$ on the string coupling $g$ will be deduced below.

The coupling of a Dbrane to background fields is given by the
Dirac--Born--Infeld action with a Chern--Simons term
\begin{equation}
        S_{p} = - \mu_{p} \int \dd^{p+1}\xi  \ee{-\Phi} \left[-\det G_{ab} +
        B_{ab} + 2\pi\al' F_{ab}\right]^{1/2}  +S_{cs}
        \label{dbi}
\end{equation}
with
\begin{equation}
        S_{cs}\equiv    i \mu_{p}\int \left(\ee{B_{ab} + 2\pi\al' F_{ab}}\wedge
        \sum_{q}C_{q}\right)_{p+1}
\end{equation}
in notation as in \cite{joe}.  The metric and Ramond-Ramond fields
produced by a source of $N$ D$p$ branes are
\begin{equation}
        \dd s^{2} = f^{{-1/2}} \dd x_{\parallel}^{2} + f^{1/2} \dd
        x_{\perp}^{2},\qquad
\end{equation}
with a dilaton
\begin{equation}
        \exp(-2\Phi) = f^{(p-3)/2},\qquad  {\rm with}\qquad f \equiv 1
        + {NgK_{p}\over 7-p} {{\al'}^{(7-p)/2}\over r^{7-p}}
\end{equation}
and an antisymmetric tensor field strength obtained from a
$p-$form potential $C_{p} \propto f^{{-1}}.$

The appearance of a massive separation mode $X$ implies that there is a
term that should be added to \refe{dbi} of the form
\begin{equation}
        \Delta S_{p} = {1\over g} \int \dd^{p+1}\xi \exp(-\Phi) \left[-\det
G_{ab}
        \right]^{1/2} {1\over 4\al'} X^{2}.
        \label{mass}
\end{equation}
Now, for an anti-Dbrane--Dbrane pair, the complete potential consists
of two terms, the potential felt by the anti-Dbrane
due to graviton-dilaton and Ramond-Ramond exchange and the
mass term for the separation mode.  The energy change if we displace
the anti-Dbrane relative to the Dbrane in the direction of the source
is
\begin{equation}
        \part_{r}f^{{-1}}\delta X (2 + {1\over 4\al'} X^{2})
        + {1\over 2\al'} f^{{-1}} X \delta X
\end{equation}
Stability requires that
the  second variation  with respect to $X $ should be positive
definite evaluated at $X=0.$
An elementary explicit calculation shows  that at a critical radius
$r_{c},$ $X$ becomes massless
when
\begin{equation}
        \left({r_{c}\over \sqrt{\al'}}\right)^{{9-p}} \approx 4(8-p)K_{p}gN
\label{critical}
\end{equation}
up to higher powers of $gN.$  At this radius, the anti-Dbrane
is no longer bound to the Dbrane, and can move away without incurring
any cost in energy.  Thus for $r<r_{c},$ anti-Dbrane--Dbrane pairs
can be nucleated
by the gradients of the background fields due to the source  D-branes.
This is {\it not} a non-perturbative
process---it follows entirely from the statement that the
D-brane--anti-D-brane configuration {\it with} a tachyon condensate is
equivalent to `nothing'\cite{sen,witten}.

Let us evaluate the value of $f$ at  this value of $r_{c}$ (\refe{critical}):
We find that $f(r_{c})=1$ up to terms of order $(gN)^{2/(9-p)}.$  Thus
the metric is close to the Minkowski metric as far as distance
measurements are concerned in the vicinity of $r_{c}.$
There is no physical significance to the value of $r_{c}$
but computing the curvature (or the gradient of the Ramond-Ramond
field strength)  at $r_{c}$\refe{critical} we find
$R_{\mu\nu}\propto 1/\alpha'.$  Thus the co\"ordinate independent
physical significance of
this critical value of the co\"ordinate $r$ is that
the curvature produced by the source Dbranes becomes of order 1
in string units.
$1/\alpha'$ appears here as a bound on curvatures produced by
Dbranes, which is quite different from  the
size of Dbranes as determined by string scattering\cite{igor}.  Our
bound is, in fact, more  reminiscent of an uncertainty relation,
since the curvature is exactly the commutator of covariant
derivatives---what we have found is that the vacuum is rendered
unstable if there are Ramond-Ramond field strength gradients that
are larger than $1/\alpha'.$  Such Ramond-Ramond field strengths
are always accompanied by curvature, though the converse does not
hold.

Since we have arrived at a conclusion that involves curvatures of order
$1/\alpha',$ while taking into account only the massless closed string
backgrounds
induced by the $N$ Dbranes, we must address the question of massive
closed string backgrounds.  Recall
that from a string field theory perspective,
massive closed string modes when integrated out lead to the
nonlinearities of supergravity.  Inclusion of massive closed string modes
is therefore related to including nonlinear effects in supergravity.
The coordinate $r_{c}$ lies in exactly the intermediate regime
between the near--horizon geometry and the asymptotic geometry.  The latter
is adequately described by single supergraviton exchange, and massive
string backgrounds are not induced in this regime.  Now, one can compute the
force felt by the anti-Dbrane in the intermediate regime at $r_c,$
and one finds that
this force is small, of order $(gN)^{1/(9-p)},$ at $r_c.$  

While pure supergravity is only accurate for $r>r_{c},$  the supergravity
calculation should still be a valid qualitative guide to
the behaviour at $r_{c},$ without
including either nonlinearities or massive closed string modes.

Notice that $r_{c}$ scales with $N$ for all $p$ as if the $N$ units of
charge were distributed uniformly over a transverse sphere of radius
$r_{c}$ (as measured in a flat metric).
This extensivity suggests that our conclusions may be valid
for $N$ small as well, though the discussion above assumes $N$ large.
Measured in a flat metric, for $p=0, N=1$ this critical radius  $r_{c}\propto
\sqrt{\al'} g^{1/9}$ should be
compared to $\ell_{11} \propto \sqrt{\al'}g^{1/3},$ the Planck length
in eleven dimensions, which appears in D0brane quantum
mechanics\cite{dfs,kp,dkps}. For small $g$ $r_{c}$ is much larger
than $\ell_{11}.$

Considering the $N$ dependence of $r_{c}$
around a configuration of $N$ D0branes, we find $r_{c}\propto N^{1/9}.$
Recall that in \cite{bfss}  such an $N$ dependence of length
scales was argued to be
required for holography on rather general grounds.  There
are no anti--D0branes in the light--front Hamiltonian\cite{bfss}, but the $N$
dependence of $r_{c}$ that we find suggests that the physics of a
finite charge radius could be incorporated in the mysterious vacuum
state of the light--front Hamiltonian.

For $p=3$ and small $gN,$ $r_{c}\propto\sqrt{\al'}(gN)^{1/6}
\gg \sqrt{\al'} (gN)^{1/4}\propto R,$ the length scale characterizing
the geometry at small values of $r.$
Thus, at small  $gN,$
we would expect the IIB string
theory to exhibit effects of a finite charge `radius'.   Since our
estimate is not valid for large $gN,$ we cannot extrapolate to this
regime, but clearly one would expect a qualitative change in the
nature of the physics when $r_{c}\approx R,$ {\it i.e.} at $gN\approx
1.$
This is also the length scale at which
stringy effects should become important
in Maldacena's conjecture\cite{malda}.   How the physics of  $r_{c}$
would appear in a sigma model with the isometry group appropriate for
the near--horizon geometry is an interesting question.

\acknowledgements
We are very
grateful to I. Klebanov, G. Lifschytz, R.C. Myers, J. Polchinski
and H. Verlinde for helpful
conversations.  H. Verlinde (unpublished) has also considered the physics of
anti--Dbrane probes in Dbrane backgrounds.
This work was supported in part by NSF grant PHY-9802484.

\end{document}